\documentclass[structabstract]{aa} 
\usepackage{graphicx}
\usepackage{float}
\usepackage{txfonts}
\usepackage{natbib}

\usepackage{graphicx} 
\usepackage{float}
\usepackage{txfonts}
\usepackage{natbib}
\usepackage{units}
\usepackage{url}

\newcommand{\lsp}{LS~I~+61$^{\circ}$303}
\newcommand{\lsi}{LS~I~+61$^{\circ}$303~}

\newcommand{\beq}{\begin{equation}}
\newcommand{\eneq}{\end{equation}}

\begin{document}

\title{Origin of the long-term modulation of radio emission of \lsi}

\author{M. Massi\inst{1} \and G. Torricelli-Ciamponi\inst{2}}
\institute{
 Max-Planck-Institut f\"ur Radioastronomie, Auf dem H\"ugel 69,
 D-53121 Bonn, Germany \\
\email{mmassi@mpifr-bonn.mpg.de}
\and
INAF - Osservatorio Astrofisico di Arcetri, L.go E. Fermi 5,
Firenze, Italy\\
\email{torricel@arcetri.astro.it}
}

\date{}

\abstract 
{One of the most unusual aspects of the X-ray binary  \lsi  is that 
at each orbit ($P_1=26.4960 \pm 0.0028$ d) one  radio outburst occurs whose  amplitude is
 modulated with $P_{\rm long}$,  a long-term period of more than 4 yr. 
It is still not clear whether the compact object of the system or the companion Be star is responsible for the long-term modulation.
} 
{We  study here the stability of  $P_{\rm long}$.
 Such a stability is expected if $P_{\rm long}$ is due to periodic ($P_2$) Doppler boosting of  
periodic ($P_1$) ejections from the accreting  compact object of the system.  On the contrary
 it is not expected if $P_{\rm long}$ is related to variations in the mass loss 
of the companion Be star.}
{We built a database of 36.8 yr of radio observations of \lsi  
	covering more than 8 long-term cycles. 
We performed timing and correlation analysis. We also compared 
the results of the analyses with the theoretical predictions  for a synchrotron 
emitting precessing ($P_2$) jet periodically ($P_1$) refilled with relativistic electrons.}
 { In addition to the two dominant
 features
at $P_1$ and $P_2$, the timing analysis  gives a
feature  at $P_{\rm long}=1628 \pm 48$ days.  
The determined value of  $P_{\rm long}$  agrees with the beat
of the two dominant  features,  i.e. $P_{\rm beat}=1/(\nu_1 -\nu_2)=1626 \pm 68$ d.
Lomb-Scargle results of radio data and model data compare very well.  The correlation 
coefficient  of the radio data oscillates at multiples of  $P_{\rm beat}$, as does the
correlation coefficient of the model data.}
{Cycles in varying  Be stars change in length and disappear after  2-3 cycles
following the well-studied case of the binary system $\zeta$ Tau.  On the contrary, in \lsi 
the long-term period is quite stable and repeats itself over the  available 8 cycles.
The long-term modulation in \lsi  accurately reflects the beat of periodical Doppler boosting (induced 
by precession) with the periodicity of the ejecta.  
The peak of the long-term modulation occurs  at the  coincidence of the maximum number of ejected particles
with the maximum Doppler boosting of their emission;  this coincidence  occurs every $\frac{1}{\nu_1 - \nu_2}$
and creates the long-term modulation observed in \lsp.}
\keywords{Radio continuum: stars - Stars: jets - Galaxies: jets - X-rays: binaries - X-rays:
  individual (\lsi) - Gamma-rays: stars}

\titlerunning{Origin of the long-term modulation of radio emission of \lsi}
\maketitle
\section{Introduction}

The system \lsp, which has an   orbital period  $P_1= \unit[26.4960 \pm 0.0028]{days}$ 
\citep{gregory02}, is composed of a compact object and
a massive star;  the star has an optical spectrum typical of a rapidly rotating
B0 V star with a decretion disc \citep{casares05}.
At each orbit one  radio outburst occurs whose  amplitude is
modulated with a long-term period  $P_{\rm long}$  of 1605$-$1667 d
\citep{martiparedes95, gregory02}. 
In this paper we  study the origin of the long-term modulation. 
 If it is attributable   to periodic Doppler boosting effects, 
due to the precession 
of a  jet associated with the compact object \citep{massitorricelli14},
then the long-term modulation has  a stable periodicity.
If it is attributable
to an identical variation in the mass loss of the Be star \citep{gregoryneish02}, then the 
long-term modulation is   rather unstable as it is  in variable Be stars \citep{rivinius13}.

After   examining the  relationship between the long-term modulation and the period
of the radio outburst (Sect. 2), we describe the precessing model (Sect. 3) and cyclic
variability in Be stars (Sect. 4). 
In Sect. 5 we perform a timing and a correlation analysis on a radio data base of
36.8 yr, that covers  more than 8  cycles of the long-term modulation. 
Finally, in Sect. 6  we discuss the implications of our results. 

\section{Relationship between the long-term modulation and the period of the radio outburst}
In this section we analyse the relationship between the long-term period $P_{\rm long}$ and the 
radio outburst period $P_{\rm average}$.
Their relationship, as shown  below, is  their mutual connection to  
the intrinsic periodicities of the system \lsp,
i.e. to the orbital period $P_1$ and to the precession $P_2$ of the jet.  

\subsection{Period of  radio outbursts $P_{\rm average}$}
\begin{figure}[]
\centering
        \includegraphics[scale=0.3, angle=-90.]{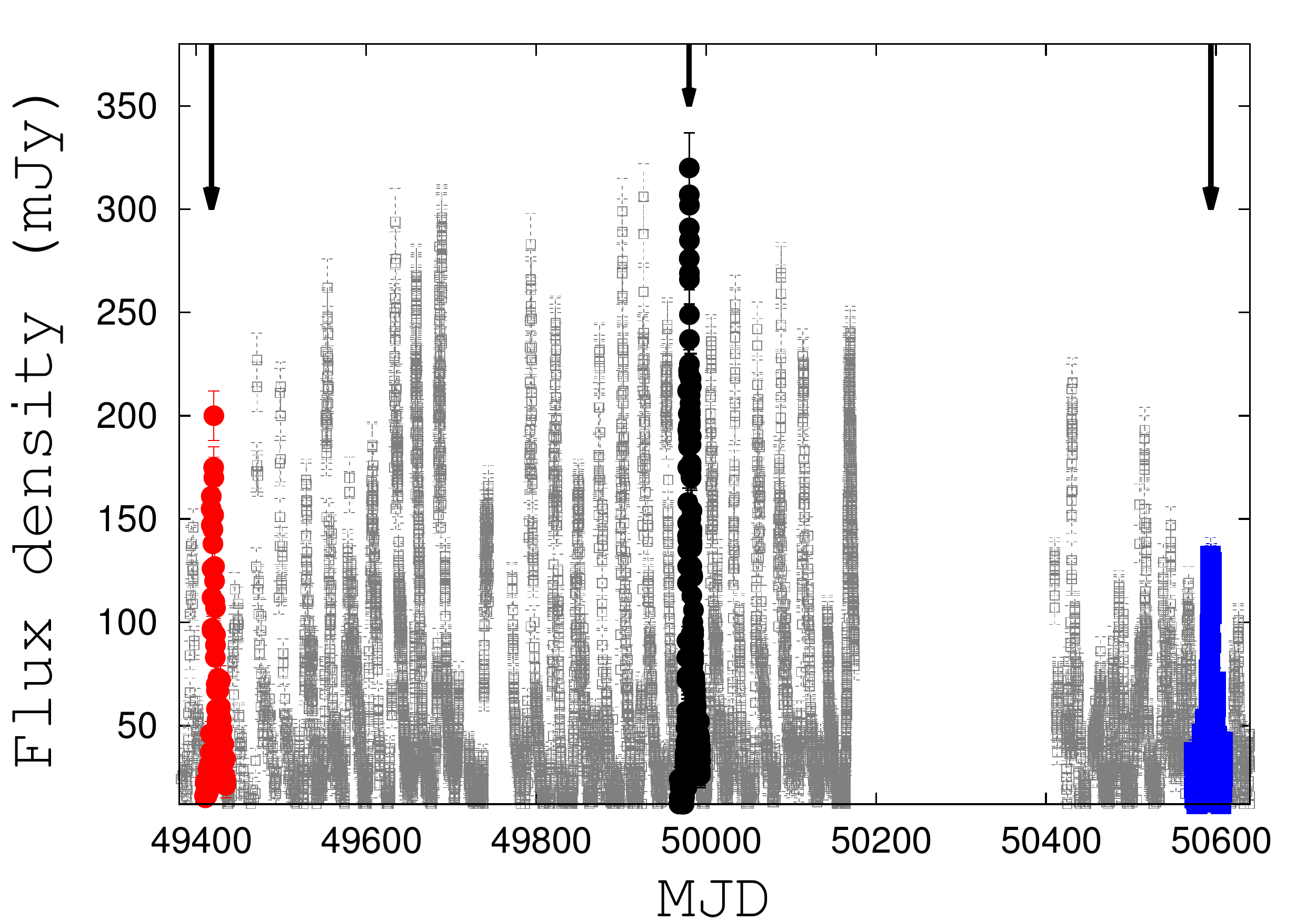}\\
        \includegraphics[scale=0.3, angle=-90.]{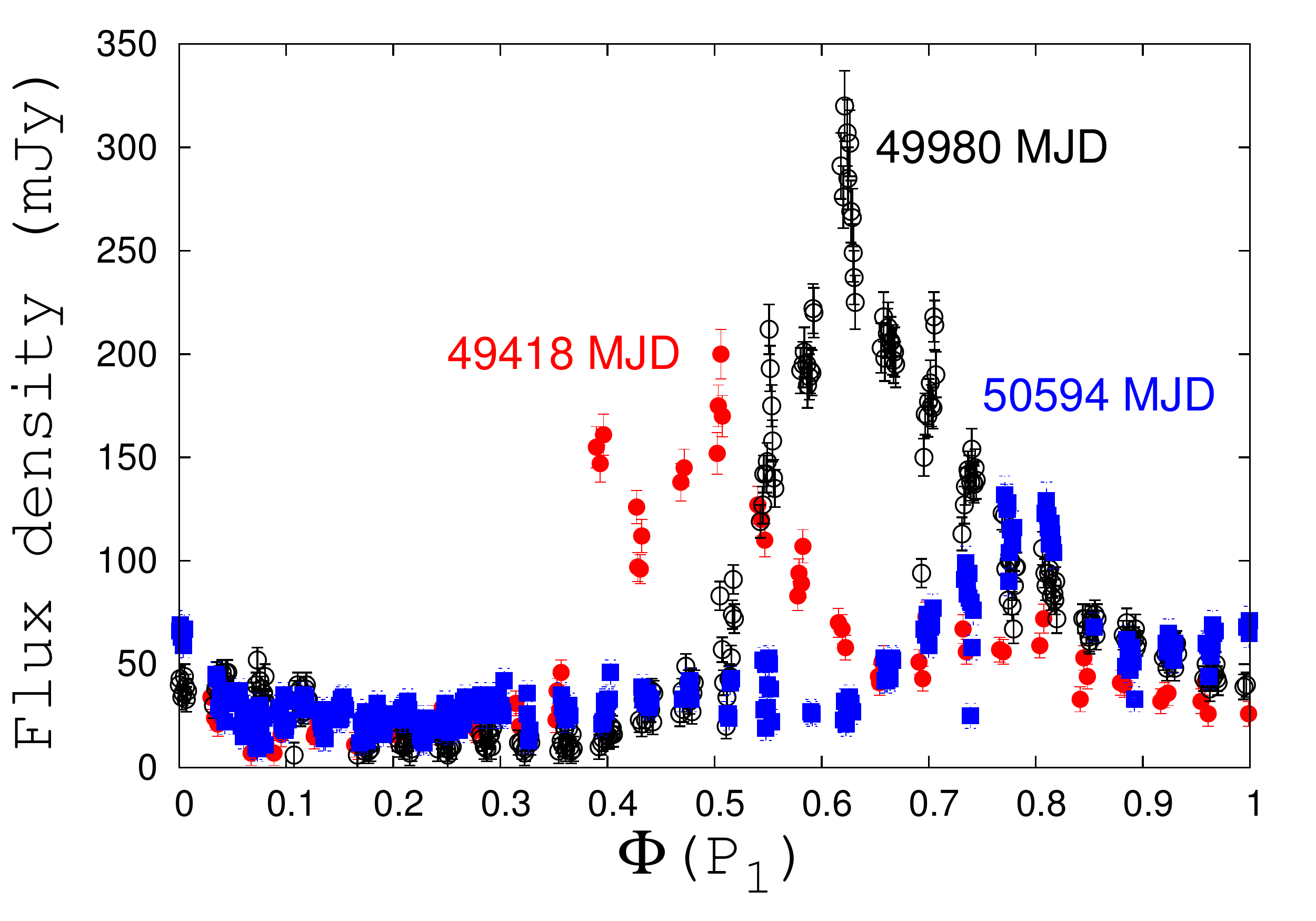}\\
        \includegraphics[scale=0.3, angle=-90.]{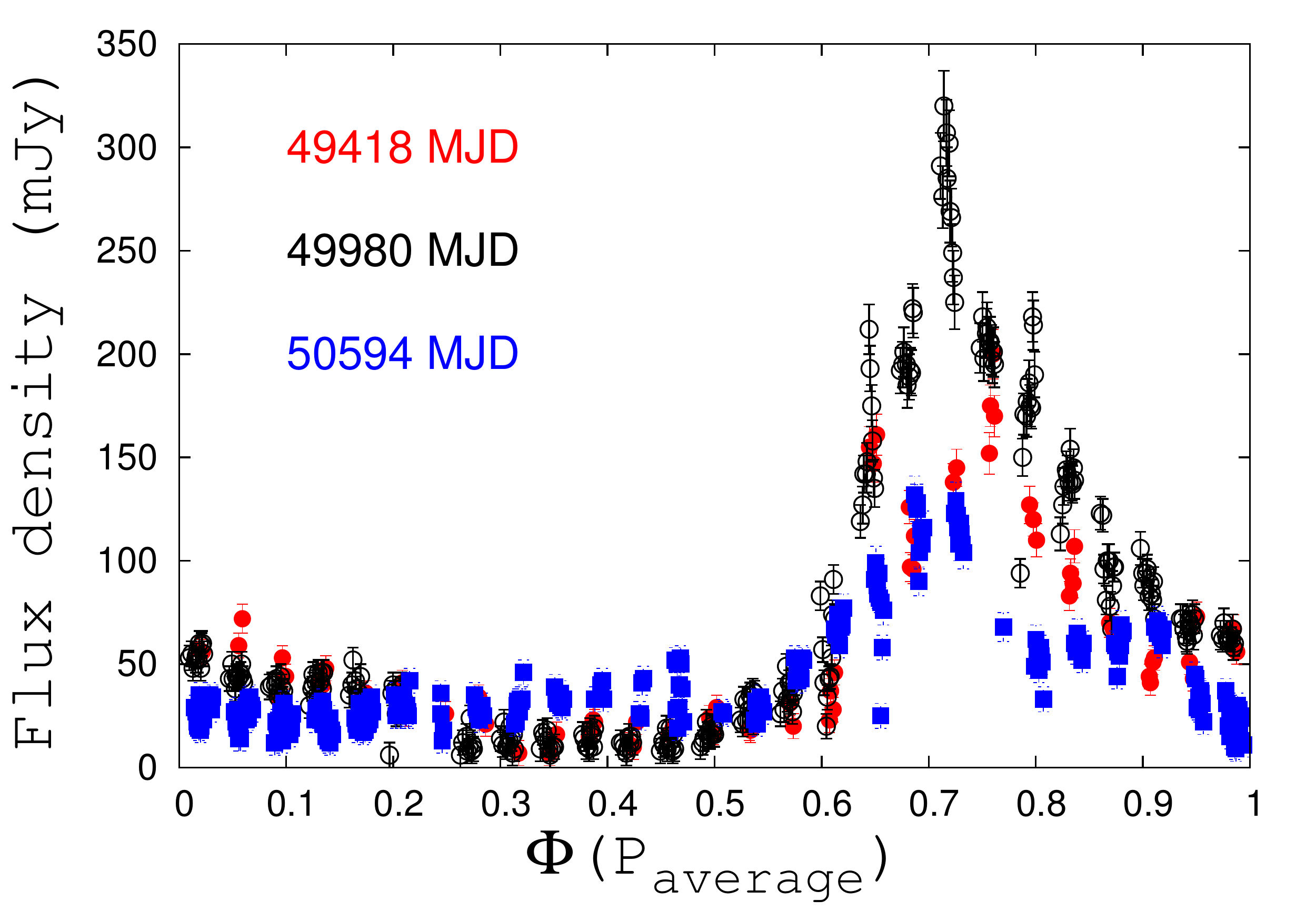}\\
\caption{Orbital phase shift in \lsp. 
Top: consecutive radio outbursts in \lsp. 
Three outbursts are indicated: one before the maximum of the long-term modulation,
one at the maximum, one towards the minimum.
Centre:
the radio light curves are folded  with 
the orbital phase $\Phi(P_1)$.
The outburst
does not follow the  orbital periodicity and shifts at  different epochs to different orbital phases. 
Bottom: the same light
curves are here folded with $P_{\rm average}$,  the periodicity
of the radio outburst \citep{jaronmassi13}.} 
\label{fig:shift}
\end{figure}

The long-term periodicity $P_{\rm long}$ 
in \lsi  is a phenomenon affecting not   only the amplitude of the radio outburst, but also
its   position along the orbit
\citep{paredes90, gregory99, gregory02}.
The variation in the position of the outburst along the orbit, i.e. the orbital phase shift,
depends on the timing residual   between the orbital period $P_1$
and the actual occurrence
of the radio outburst. 
The radio outburst occurs  {nearly} but not exactly at $P_1$.
\citet{gregory99} showed that when radio outbursts
are predicted to occur with $P_1$ then  there are  timing residuals of up to several days  
that follow a  sawtooth trend with   period $P_{\rm long}$.
The same  phenomenon has been analysed in terms of orbital phase shift by \citet{paredes90}.
An example is  given in Fig. \ref{fig:shift}.
The top panel shows one outburst before the maximum
of the long-term modulation at  49418 MJD (red), one outburst during the maximum at  49980  MJD (black),
and one outburst towards the minimum at  50594 MJD (blue).
In Fig.  \ref{fig:shift}-centre we plot the three outbursts as function
of the  orbital phase $\Phi$,  
\beq
\Phi(P_1)={(t-t_0)\over P_1} -int \Big [{(t-t_0)\over P_1} \Big ] 
,\eneq
where $t_0$=43366.275 MJD \citep{gregory02}. 
The figure shows that the first outburst peaks at phase 
$\Phi (P_1) \sim 0.5$, whereas  
the second outburst is  shifted at $\Phi \sim 0.6,$ and the third  
at $\Phi \sim 0.8$.
As can be seen in both Fig. \ref{fig:shift}-top and centre,  
 there is a variation in the amplitude of the outburst at the same time as the orbital phase shift.
The period of the amplitude modulation and the period of the sawtooth function are the same 
 \citep{gregory99,gregory02}, so clearly the origin of the two phenomena  are  also  the same.
Why is there  an orbital phase shift? or better, Why is there  a timing residual with respect 
to the orbital period $P_1$?
A saw-tooth function in the residuals is  indicative of a systematic error,
otherwise residuals should be randomly  distributed.
An intuitive answer is that $P_1$ is not the periodicity of the radio outburst.
The first group to notice that the radio outburst   periodicity  is not equal to the orbital periodicity  is \citet{ray97}.
In 1997 they  reported
a period for the radio outburst of $\unit[26.69 \pm 0.02]{d}$ and discuss how
their best estimate of the period is significantly different
($9\sigma$) from the best previously published value of 
$\unit[26.496 \pm 0.008]{d}$  \citep{taylorgregory84}.
Based on the analysis of 6.5 years of Green Bank Interferometer (GBI) radio data, the period of 
$P_{\rm average} = \unit[26.704 \pm 0.004]{d}$
for the radio outbursts in \lsi 
has been confirmed, and  
noise-limited timing residuals between radio outbursts and predicted outbursts 
at $P_{\rm average}$ have been determined \citep{jaronmassi13}.
The bottom panel of Fig. 1 shows that the three outbursts at the three
different epochs all occur at the same $\Phi(P_{\rm average})$.
 The reason  why the outburst period is
called $P_{\rm average}$ is explained in the following section.
\begin{figure*}[]
\centering
\includegraphics[scale=0.6, angle=-90.]{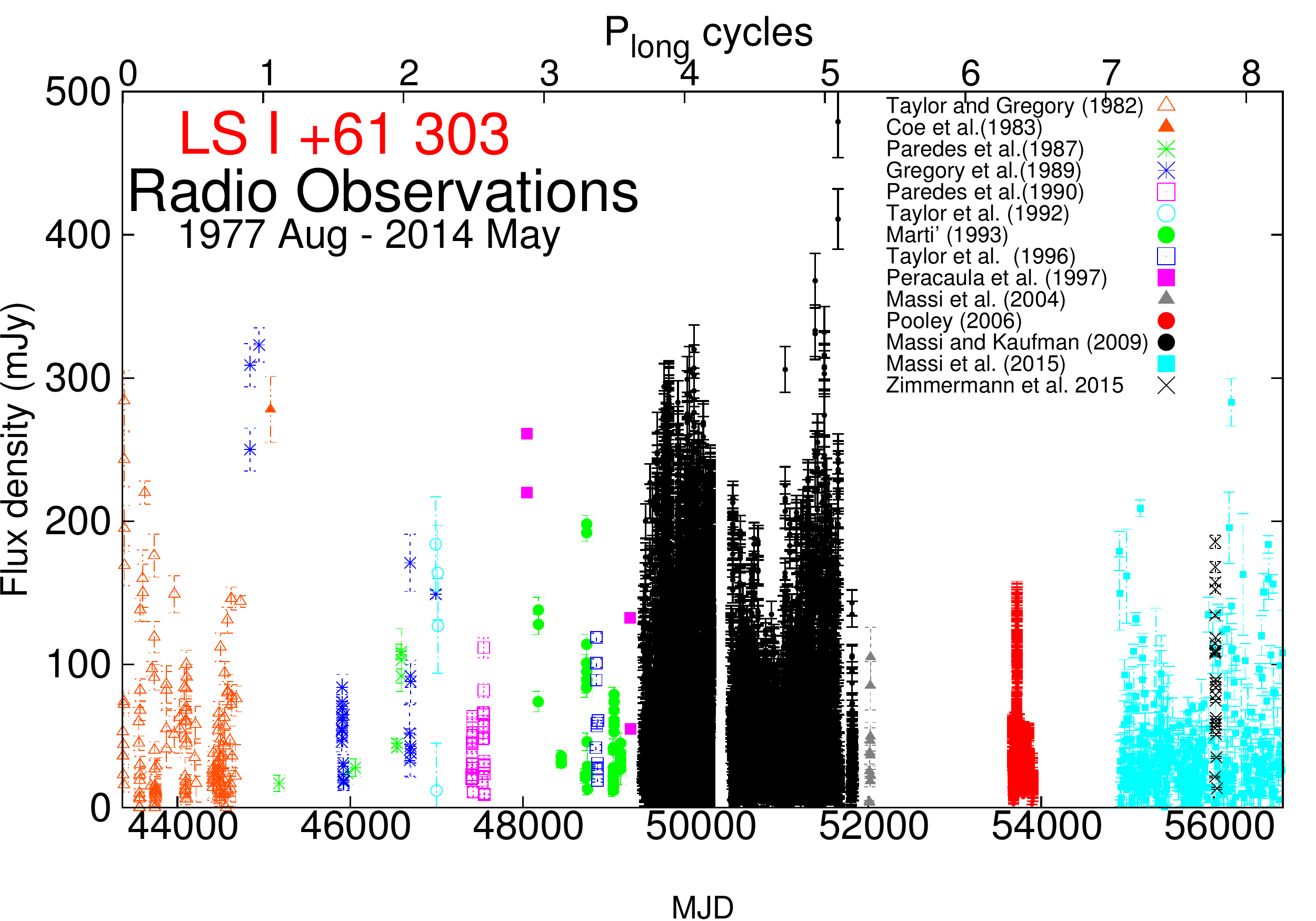}\\
\caption{Radio data of \lsi at 5, 8, 10, and 15 GHz from the literature.
At the top x-axis are cycles of the long-term modulation, i.e.
$t - t_0 / 1626$ 
(see Sect. 4). The archive covers more than 8 cycles.} 
\label{fig:alldata}
\end{figure*}
\subsection{Beat between $P_1$ and $P_2$}
Another peculiarity  of the radio emission from this system
is the variability of its morphology.
Radio images at high angular resolution  show 
an elongated feature that changes from a double-sided to a one-sided morphology and 
that continuosly changes position angle as well. A precession
period 
of 27--28 d was estimated by VLBA astrometry   \citep{massi12}. 
The observed flux density ($S_o$) of a synchrotron source is the intrinsic
flux density ($S_i$) times the Doppler boosting factor (DB), which  depends
on the  ejection angle. 
A periodic change in ejection angle of a radio emitting source, because of precession,
changes  the  Doppler factor periodically; as a consequence, a periodicity should be observed
in the flux density.
Timing analysis of GBI radio data has revealed two periods \citep{massijaron13}, the already
known orbital period 
$P_1 = \unit[26.49 \pm 0.07]{d}$ ($\nu_1=\unit[0.03775]{d^{-1}}$) and
a new period  $P_2 = \unit[26.92 \pm 0.07]{d}$ ($\nu_2 = \unit[0.03715]{d^{-1}}$)
consistent with the precessional period of 27--28 d
indicated by the VLBA astrometry \citep{massi12}. 
The orbital period can result in a timing analysis  if there is a periodical
variation in the intrinsic flux density ($S_i$).
The hypothesis that  the compact object in \lsp, which accretes  material from the  Be wind,
undergoes a  periodical  ($P_1$) increase in the accretion  rate $\dot{M}$ at a  particular orbital phase
along an eccentric orbit
has been suggested and developed  by several authors  \citep{taylor92, martiparedes95, boschramon06, romero07}.
Recently, the presence of $P_1$ and $P_2$  have been confirmed as stable features
in more recent  Fermi-LAT and OVRO  monitorings of the  GeV gamma-ray emission \citep{jaronmassi14} 
and of the radio emission \citep{massi15}, respectively. 

The aim of the timing analysis was to obtain a better determination of the precessional period, but there was
an additional and  totally unexpected result:  when combining $\nu_1$ and $\nu_2$  
      one obtains 
    $P_{\rm average} =    \frac{2}{\nu_1 + \nu_2} =
  \unit[26.70 \pm 0.05]{d}$  and 
  $P_{\rm beat}={1\over\nu_1-\nu_2}=
\unit[1667 \pm 393]{d}$  \citep{massijaron13}.
 In other words, 
   the beat of  $P_1$ and $P_2$ creates $P_{\rm beat}$ (i.e. $P_{\rm long}$)  that modulates  their average
$P_{\rm average}$.
\lsi seems to be  one more case   in astronomy of a ``beat''.
The first astronomical case was  a class  of Cepheids, afterwards called
Beat-Cepheids \citep{oosterhoff57}.
This  phenomenon occurs when two physical processes create
stable variations  of  nearly equal frequencies ($\nu_1$,$\nu_2$).  
The very small difference in frequency
creates a long term variation ($\nu_1-\nu_2$) modulating their average ${(\nu_1 + \nu_2)}/2$.
\section{Precessing jet  model}
In our  previous paper \citep{massitorricelli14},  
we linked the two periodicities $P_1$ and $P_2$ to two physical
processes: the periodic ($P_1$) increase  of relativistic electrons
in a conical  jet due to   $\dot{M}$  variations
and  the periodic ($P_2$)  Doppler boosting of the
emitted radiation by relativistic electrons due to  jet precession.
We have shown that 
the synchrotron emission of such a jet,
calculated  and fitted   to the GBI observations,  
reproduces {both}  the long-term  modulation {and} 
the orbital phase shift of the radio outburst \citep{massitorricelli14}.
The peak of the long-term modulation occurs when the jet electron density is around its maximum  and
the approaching jet forms the smallest possible angle with the line of sight.
This coincidence of the maximum number of emitting particles and  the maximum Doppler boosting of their emission
occurs every $\frac{1}{\nu_1 - \nu_2}$
and  creates the long-term modulation observed in \lsp.
In this context, if  
no significant variations in $P_1$ and $P_2$ occur in the time interval
under examination, then in that interval the long-term modulation 
 remains stable.

\section{Cyclic variability in the disc of some Be stars}
The known long-term variability in Be stars is that of the violet-to-red cycles 
(V/R variations) in which the two peaks of the emission lines
vary in height against each other. 
The variations correspond to an evolution of the disc itself
where the disc undergoes a global one-armed oscillation instability 
that manifests itself in the V/R variability. This instability (progressing density wave) 
may  eventually lead to the disruption of the disc, which after years or decades fills 
again \citep[][and reference therein]{grudzinka15}, and to 
cycle lengths that  are not constant but vary from cycle to cycle 
\citep[][and references therein]{rivinius13}. 
We examine in detail the well-studied  case of $\zeta$ Tau, which is also a Be star in a  binary system.
Over the last 100 years $\zeta$ Tau has gone through active stages  characterized by pronounced long-term variations and quiet stages. 
From 1920 to about 1950 there were no  variations. A very clear variability started 
afterwards  and lasted until 1980 with  2 cycles of 2290 days and 1290 days.
From 1980 to 1990 the star was again stable. Around 1990 the disc  again entered  into 
an active stage  and  3 consecutive cycles of 1419 d,
1527 d, and 1230 d were determined \citep[][and references therein]{stefl09}.
In this context,  radio variations induced by variations in the Be disc  
are predicted to be rather unstable, i.e. to have cycles of different lengths and,  
at some epochs, to disappear.

\section{Radio data base of \lsp}

\begin{figure*}[htbp]
\centering
        \includegraphics[width=7cm, height=14cm, angle=-90.]{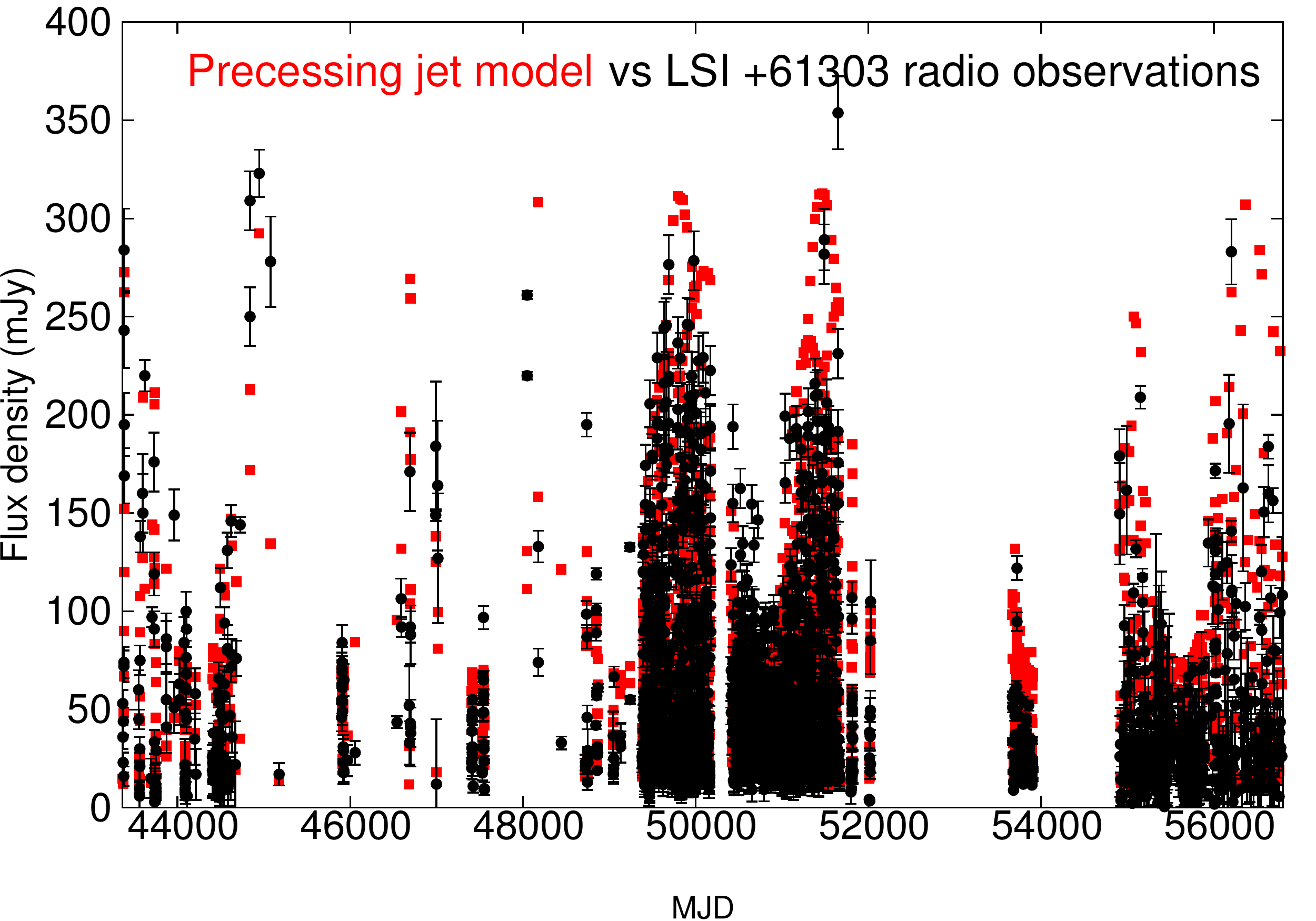}\\
\caption{ Model data (red) and radio observations  (black) of \lsi averaged over one day (Sect. 5).}
\label{fig:model}
       \includegraphics[width=8cm, height=8cm, angle=-90.]{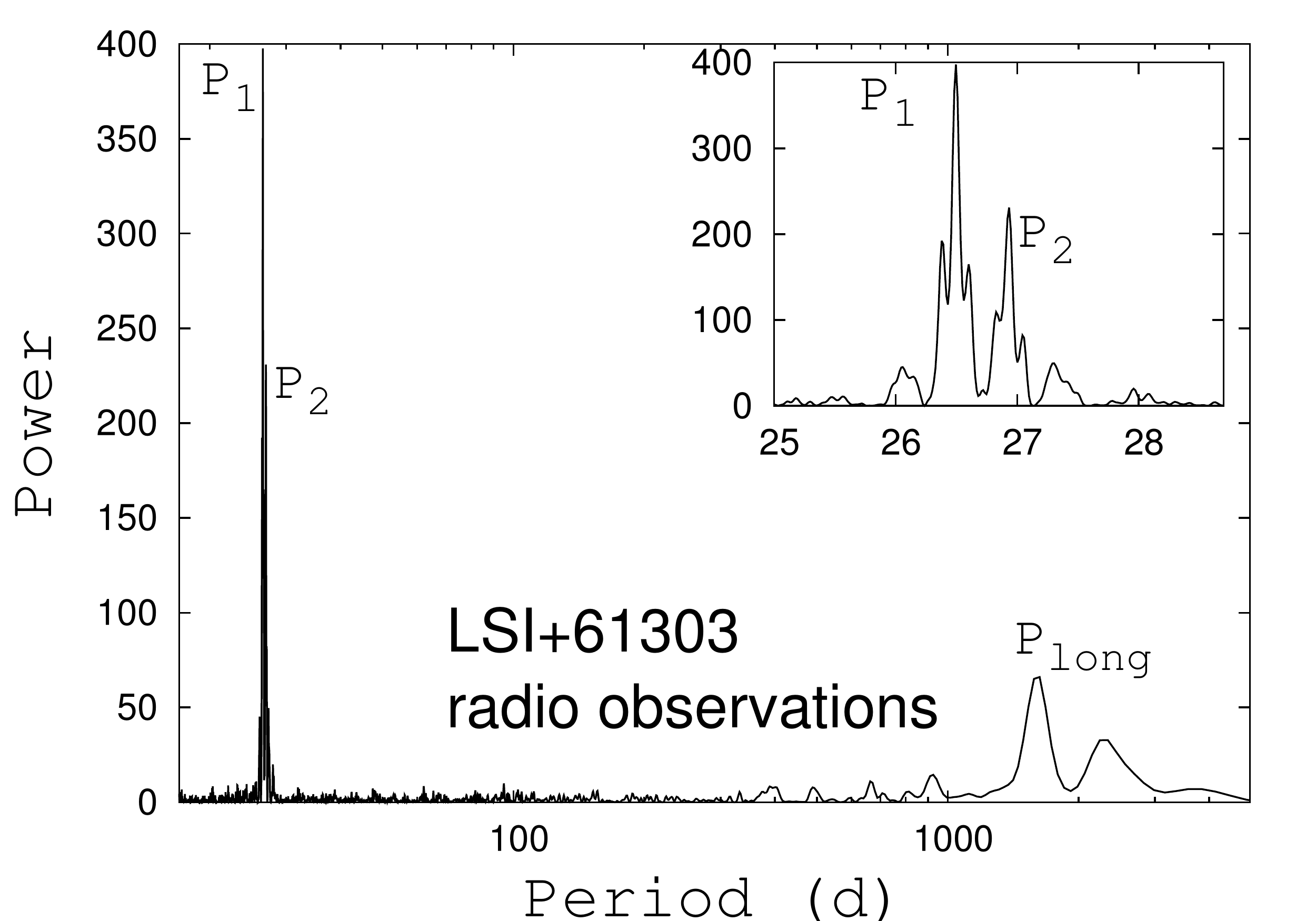}
        \includegraphics[width=8cm, height=8cm, angle=-90.]{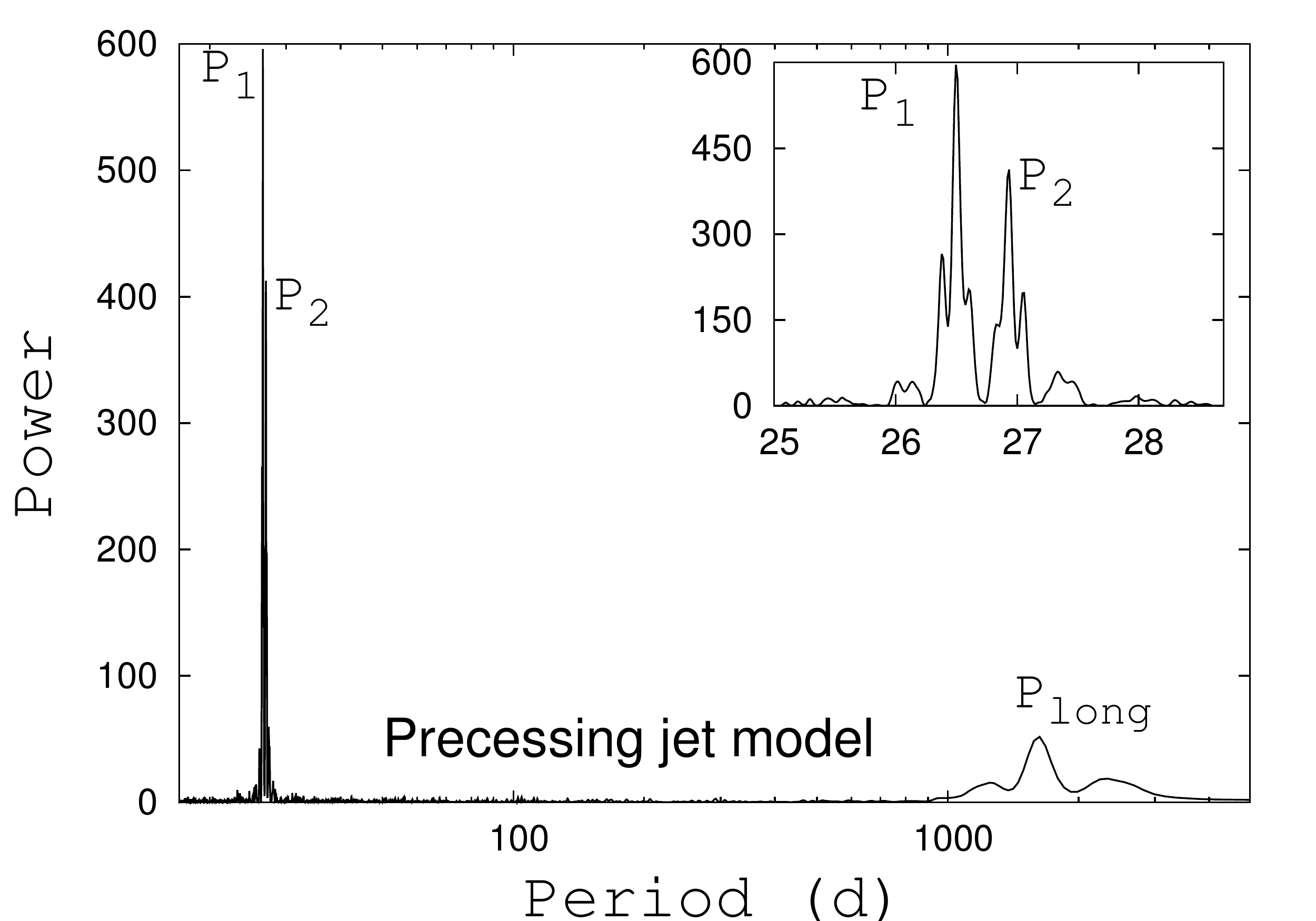}
\caption{Lomb-Scargle timing analysis of the observations  (left) and the model data
(right) of Fig. \ref{fig:model}.}
\label{fig:scargle}
\includegraphics[width=8cm, height=8cm, angle=-90.]{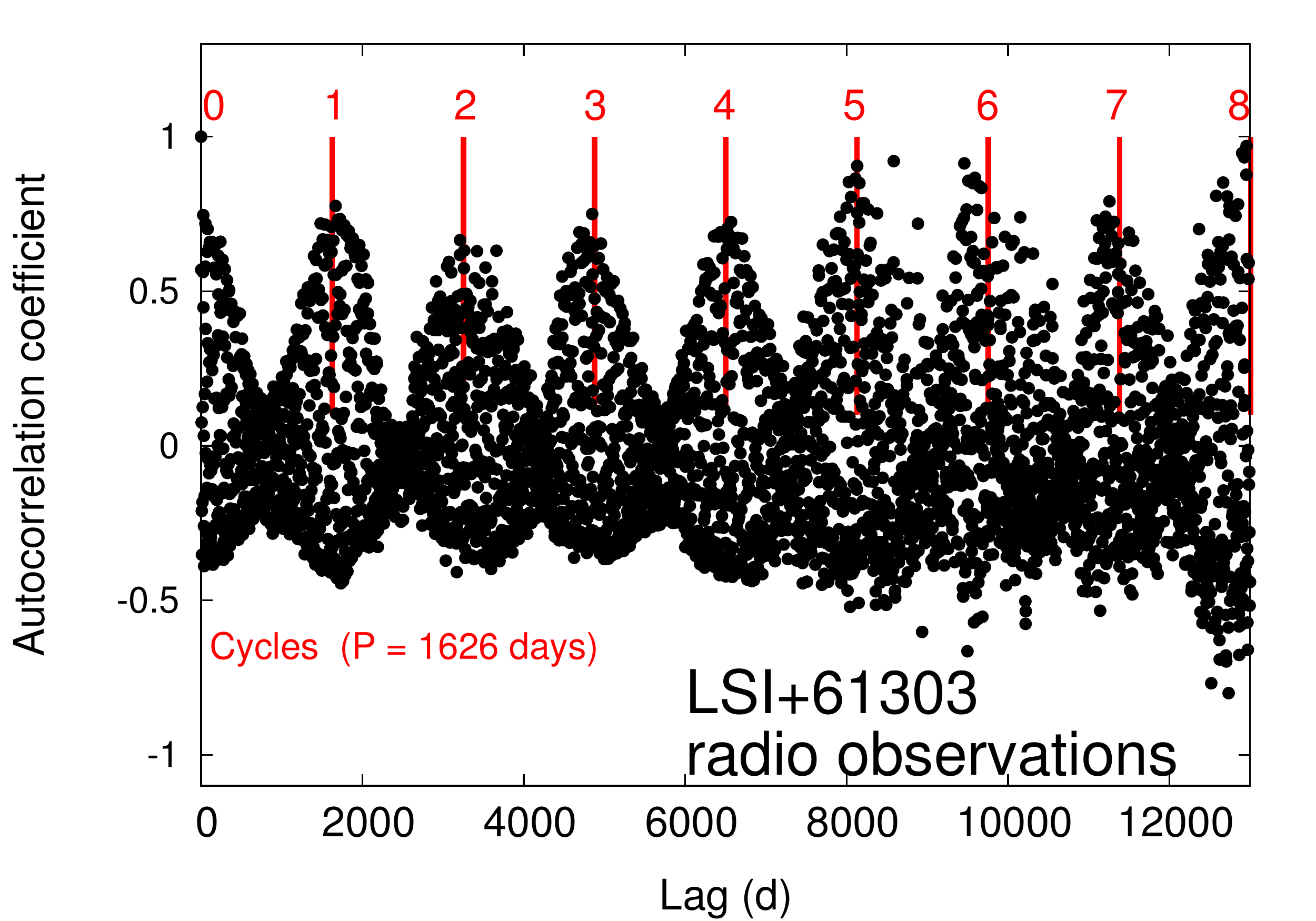}
        \includegraphics[width=8cm, height=8cm, angle=-90.]{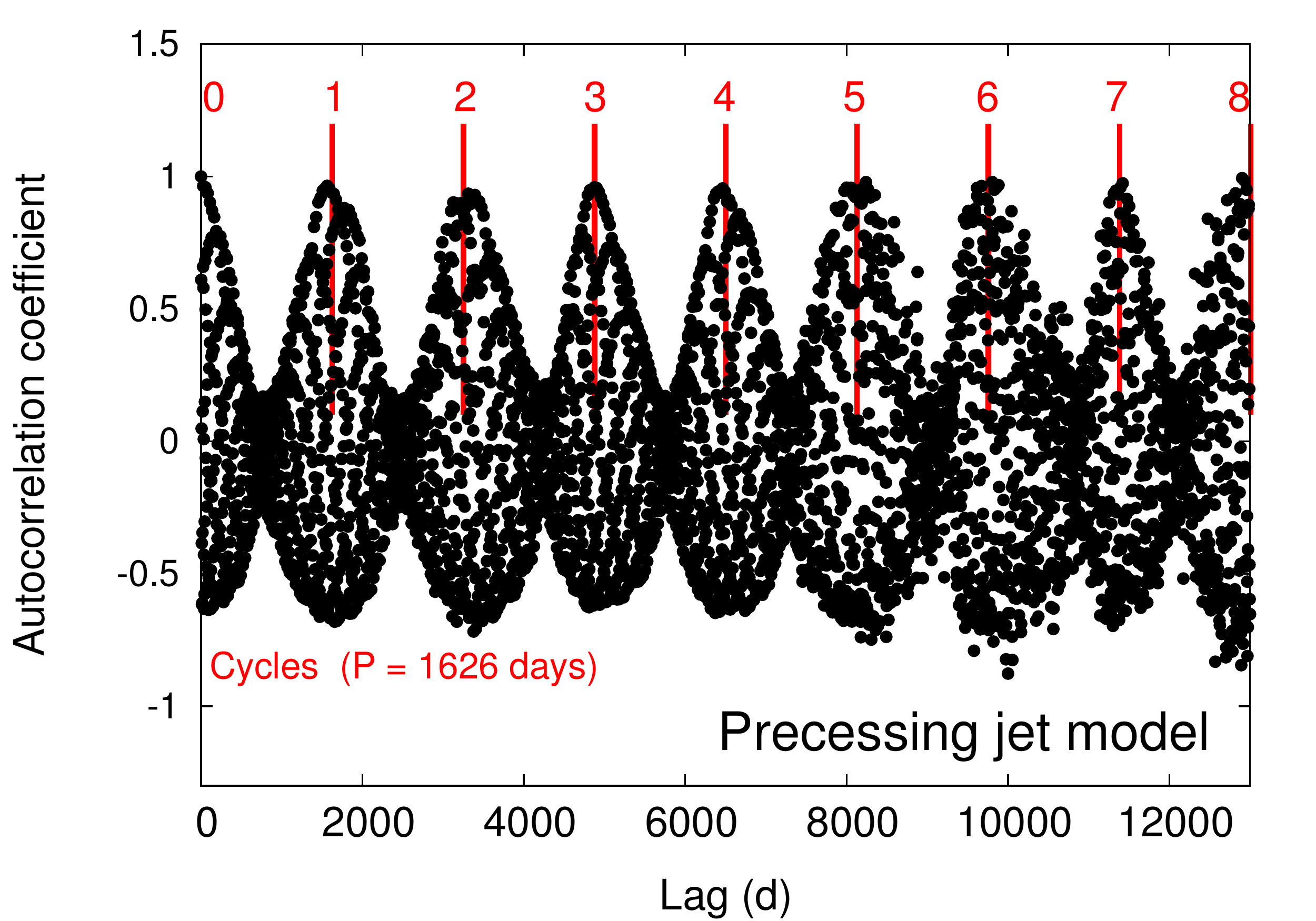}
\caption{Correlation coefficient vs time for  observations  (left) and  model data
(right), both averaged over three days.}
\label{fig:correl}
\end{figure*}

We collected data of \lsi from the literature at 5, 8, 10, and 15 GHz (see Appendix)
covering the interval
43367 MJD $-$ 56795 MJD, i.e. 36.8 yr.
To date, this is the largest archive   on \lsp.
The data are shown in Figure \ref{fig:alldata}. 
The top x-axis shows the number of cycles for a long-term period of 1626 d (see below): 
the data set covers 8 long-term cycles.
In Fig. \ref{fig:model} we show the data  overlapped with our physical model
of the precessing jet 
assuming the same sampling and $\nu=8$ GHz.
Our  model was fitted previously  \citep{massitorricelli14}
on 6.7 yr of radio data (49380 MJD -  51823 MJD). As we show
in Fig. 3, the model also reproduces well  36.8 yr of data. 

 In
order to search for possible periodicities we used the Lomb-Scargle
method, which is very  efficient  on irregularly sampled data
\citep{lomb76, scargle82}. We use the algorithms of the UK Starlink software
package, PERIOD (\url{http://www.starlink.rl.ac.uk/}).
The statistical significance of a period is calculated in PERIOD
following the method of Fisher randomization as outlined in
\citet{nemec85}.
 One thousand randomized time series are formed and the
periodograms are calculated.
The proportion of permutations  in a given
frequency window that give a peak power higher than that
of the original time series provides an estimate of $p$, the
probability that there is no such  periodic component.
A derived period is considered
significant for $p < 0.01$
 and  marginally significant  for $0.01
< p < 0.10$ \citep{nemec85}.

The spectrum is shown in Fig.4.
We have two dominant features at 
 $P_1 = \unit[26.496 \pm
  0.013]{d}$  and $P_2 = \unit[26.935
  \pm 0.013]{d}$,  evident in the zoom in the upper right panel.
Period $P_1$ coincides with the orbital period \citep{gregory02} 
and period $P_2$ with the precession period \citep{massijaron13, massi15}.
In the spectrum there is a third feature at $P_{\rm long}=1628 \pm 49$ d.
For all  periods the randomization test gives $0.00 < p < 0.01$. 
It is interesting to compare this spectrum with that of the model shown in Fig. 4-right.
Periods $P_1$, $P_2$, and $P_{\rm long}$ are all there. Period $P_{\rm long}$ shows two side lobes
at 1265 d and at 2335 d; 
 the  small feature to the right of $P_{\rm long}$ in Fig. 4-left corresponds to 2335 d.

By comparing the light curves in Fig. \ref{fig:model}  and the spectra, i.e. 
  Fig. \ref{fig:scargle}-left  with Fig. \ref{fig:scargle}-right, 
it is clear   that  the precessing jet model  
able to reproduce the light curve of 36.8 yr of radio data of \lsi 
also  reproduces well  its Lomb-Scargle spectrum.
The most important result from the analysis of the largest archive of  data 
available until now on \lsi  is that
it is evident for the first time that $P_{\rm long}$
  is not a main characteristic of the Lomb-Scargle spectrum; 
the main characteristics are $P_1$ and $P_2$.
We can compute the beating of the found $P_1$ and $P_2$ and check how 
$P_{\rm beat}$ compares with $P_{\rm long}$. The result is $P_{\rm beat}=(P_1 P_2)/(P_2-P_1)=
1626 \pm 68$ d, clearly in agreement with $P_{\rm long}=1628 \pm 49$ d. 

How stable is the found $P_{\rm long}$   over  the 8 cycles? This is the fundamental question
 of this work. A  powerful method for studying the stability of a periodic signal over time
is its autocorrelation. If a period is stable the correlation coefficient should
appear as an oscillatory sequence with peaks at multiples of that 
period.
We first examine the correlation coefficient for model data. 
The result is shown in Fig. \ref{fig:correl}-right. In the Lomb-Scargle spectrum,
$P_{\rm long}$--
given by the beating of $P_1$ and $P_2$ -- 
is a minor feature; however,  it becomes  apparent in the plot of the correlation coefficient. 
We draw lines at multiples of $P_{\rm beat}=1626$.
Since $P_1$ and $P_2$ are constant in our model, it is clear that the model predicts
a stable repetition of their beating frequency. 
We now compare the correlation of model data with the correlation
of the observations. The correlation coefficient
is shown in Fig. \ref{fig:correl}-left.    
Except for a slight drop in the correlation in the first cycles,
the correlation coefficient of \lsi observations compares well with the
correlation coefficient of the model data. 
The  small drop in correlation is likely attributable
to the data at 5 GHz present in the first 3.7 cycles (see Appendix).
The scattering of the correlation coefficient
around cycle 6 is  due to the  lack of data around that cycle (see Fig. \ref{fig:alldata});
  scattering is also present  in the 
correlation of model data (Fig. \ref{fig:correl}-right).

\section{Discussion and conclusions}

We built a radio archive of 36.8 yr for the source \lsi
and performed timing and correlation analysis.
Our conclusions  are the following:
\begin{enumerate}
\item
In the system \lsi  a  periodic radio outburst is observed of  
 $P_{\rm average}=26.704 \pm 0.004$ d modulated by a long-term period of $1628 \pm 49$ d.
However, the Lomb-Scargle spectrum is  dominated by  two other periodicities, 
 $P_1 = 26.496 \pm 0.013$ d  and $P_2 = 26.935 \pm 0.013$ d,
and the observed periodicities $P_{\rm average}$ and $P_{\rm long}$
are equal to the beat of $P_1$ and $P_2$.

\item
Synchrotron emission from a precessing jet model, with precession period $P_2$,
regularly  (with period $P_1$) 
refilled with relativistic electrons, reproduces the observed    
36.8 yr light curve of \lsi and also its Lomb-Scargle spectrum. 
\item
The correlation coefficient of the observations shows a regular oscillation, 
as does the correlation coefficient of the model data 
with peaks at multiple of  $P_{\rm long}$.  
\end{enumerate}

The period $P_{\rm long}=1628\pm 49$ d is  stable over 8 cycles.
This is not what is expected from Be wind variations, which   are rather unstable
and change from cycle to cycle or  even disappear. 
A stable $P_{\rm long}$ is  what is expected 
for a beat of periods  $P_1$ (orbital period)
and $P_2$ (precession) with no significant variations during the 8 examined cycles.
The peak of the long-term modulation occurs when the jet electron density is around 
its maximum  and the approaching jet  forms  the smallest possible angle 
with the line of sight.
This coincidence of the maximum number of emitting particles and  the maximum Doppler 
boosting of their emission occurs every $\frac{1}{\nu_1 - \nu_2}$
and  creates the long-term modulation observed in the radio emission of \lsp.

Finally, periods $P_1$ and $P_2$ have also been determined  in the Lomb-Scargle spectrum
of gamma-ray emission at apastron of Fermi-LAT data \citep{jaronmassi14}.
Moreover,  \citet{paredesfortuny15} 
  discovered that the orbital phase shift also affects  the variations in the
equivalent width (EW) of the H$\alpha$ emission line from \lsp.
The orbital  phase  shift
 (Sect. 2.1) suggests that  $P_{\rm average}$ is the true period 
for the EW variations. 
The EW variations  show  a long-term modulation as well \citep{zamanov13}.
The presence of both long-term period and $P_{\rm average}$, as well as  
the direct detection of $P_1$ and $P_2$, indicates therefore
that  the precessing jet also induces    variations in 
EW(H$\alpha$) and gamma-ray emission.
 
\begin{acknowledgements}

We are grateful to Hovatta Talvikky for providing the OVRO data and to
Guy Pooley for providing the Ryle data.
We  acknowledge several  helpful discussions with Frederic Jaron and J\"urgen Neidh\"oferr.
We acknowledge Atsuo Okazaki for suggesting the case of the Be star  $\zeta$ Tau.
We would like to thank the anonymous referee for the helpful comments.
The OVRO 40 m Telescope Monitoring Program is supported
by NASA under awards NNX08AW31G and NNX11A043G, and by the NSF
under awards AST-0808050 and AST-1109911. 
This research is based on observations with the 100 m telescope of the MPIfR (Max-Planck-Institut für Radioastronomie) at Effelsberg.
The Green Bank Interferometer is a facility of the National
Science Foundation operated by the NRAO in support of NASA High Energy
Astrophysics programs.
\end{acknowledgements}

\bibliographystyle{aa}

\begin{appendix}
        \section{Selection of frequencies for the radio data base}
We selected from the literature data at  5, 8, 10,  and 15 GHz.
To reduce inhomogeneities in the archive, we excluded   frequencies lower 
than 5 GHz because of the high flux that
the `second peak' can have at low frequency. 
The  
periodic radio outburst in \lsi is in fact not a simple, one-peak outburst. 
It may have  a rather complex structure, mainly  dominated by two consecutive peaks.
These two peaks are 5-8 days apart and have different spectral characteristics;
the  first peak  has a flat/inverted spectrum and 
the second outburst  is optically thin, i.e. it dominates at low frequencies. 
At 8 GHz the second peak is often a minor peak, hardly traceable; instead, 
simultaneous observations at 2 GHz show a strong  second peak, even 
 larger at some epochs than the first peak
\citep{massikaufman09, massi14, zimmermann15}.

We included data at 5 GHz to be able to cover the first cycles. 
However, at this frequency the second peak still seems to
be  significant and biases  the correlation analysis slightly.
The small drop in correlation coefficient from 1 to 0.7$-$0.8 (Fig.5-left)
 is  likely attributable to the  second peak 
at 5 GHz; for example 
in Table 1 in \citet{taylorgregory82}, a  second
peak  of 91 $\pm$ 9mJy can be seen at 2444106.95 JD following the first peak (six  days earlier)   of 100 $\pm$ 10 mJy.
\end{appendix}
\end{document}